\documentclass[preprint,11pt]{elsarticle}
\biboptions{sort&compress}

\usepackage{amsmath,amssymb,amsfonts}%
\usepackage{xcolor}%
\usepackage{algorithm}%
\usepackage{algorithmicx}%
\usepackage{algpseudocode}%
\usepackage{listings}%
\usepackage{enumitem}
\setlist[enumerate,1]{label=\arabic*.}
\setlist[enumerate,2]{label=\roman*)}

\usepackage{graphicx}%
\usepackage{tikz}
\usepackage{multirow}%
\usepackage{mathtools}
\usepackage{mathrsfs}%
\usepackage{bm}
\usepackage{nicefrac}
\usepackage[figuresright]{rotating}%
\usepackage[title]{appendix}%
\usepackage{textcomp}%
\usepackage{subcaption}
\setlist[enumerate,1]{label=(\arabic*)}
\setlist[enumerate,2]{label=\roman*)}

\newcommand{\sty}[1]{\mbox{\boldmath $#1$}}

\newcommand{\fu}{\sty{ u}}
\newcommand{\fz}{\bold z}

\newcommand{\fC}{\sty{ C}}

\newcommand{\fx}{\bold x}
\newcommand{\ff}{\sty{ f}}
\newcommand{\ft}{\sty{ t}}
\newcommand{\fg}{\sty{ g}}
\newcommand{\fn}{\sty{ n}}
\newcommand{\fy}{\bold y}
\newcommand{\ftheta}{\boldsymbol{\theta}}

\newcommand{\feps}{\mbox{\boldmath $\varepsilon $}}
\newcommand{\fsig}{\mbox{\boldmath $\sigma$}}

\newcommand{\argmin}{\mathop{\mathrm{arg\,min}}}

\DeclareFontFamily{U}{mathx}{\hyphenchar\font45}
\DeclareFontShape{U}{mathx}{m}{n}{
	<5> <6> <7> <8> <9> <10>
	<10.95> <12> <14.4> <17.28> <20.74> <24.88>
	mathx10
}{}
\DeclareSymbolFont{mathx}{U}{mathx}{m}{n}
\DeclareMathSymbol{\bigtimes}{1}{mathx}{"91}
\algnewcommand\algorithmicbreak{\textbf{break}} 
\algnewcommand\Break{\algorithmicbreak{} }%
\algnewcommand\algorithmicdata{\textbf{Data:}}
\algnewcommand\Data{\item[\algorithmicdata{}]}%
\algrenewcommand\algorithmicrequire{\textbf{Input:}}
\algrenewcommand\algorithmicensure{\textbf{Output:}}

\usepackage{scalerel}
\newlength\bshft
\def\fakebold#1{\ThisStyle{\ooalign{$\SavedStyle#1$\cr%
			\kern-\bshft$\SavedStyle#1$\cr%
			\kern\bshft$\SavedStyle#1$}}}

\journal{Computer Methods in Applied Mechanics and Engineering }

\begin{document}

\begin{frontmatter}

\title{A physics-informed GAN Framework based on Model-free Data-Driven Computational Mechanics}

\author[rub]{Kerem Ciftci\corref{cor1}}
\ead{Kerem.Ciftci@rub.de}
\author[rub]{Klaus Hackl}
\ead{Klaus.Hackl@rub.de}
\cortext[cor1]{Corresponding author}
\address[rub]{Institute of Mechanics of Materials, Ruhr University Bochum, Universit\"atsstrasse 150, 44801 Bochum, Germany.}

\begin{abstract}
	Model-free data-driven computational mechanics, first proposed by Kirchdoerfer and Ortiz, replace phenomenological models with numerical simulations based on sample data sets in strain-stress space. In this study, we integrate this paradigm within physics-informed generative adversarial networks (GANs). We enhance the conventional physics-informed neural network framework by implementing the principles of data-driven computational mechanics into GANs. Specifically, the generator is informed by physical constraints, while the discriminator utilizes the closest strain-stress data to discern the authenticity of the generator's output. This combined approach presents a new formalism to harness data-driven mechanics and deep learning to simulate and predict mechanical behaviors.
\end{abstract}


\end{frontmatter}

\section{Introduction}\label{sec:intro}
The simulation of boundary value problems typically contains two equations: conservation and constitutive laws. While conservation laws are derived from universal principles, constitutive laws are usually obtained by fitting model parameters to given strain-stress data \cite{timoshenko:1983}. Nevertheless, material modeling can be ill-posed and adds uncertainties to the solutions, particularly in highly complex systems. The model-free data-driven method, introduced by Kirchdoerfer and Ortiz \cite{kirchdoerfer:2016}, bypasses the step of material modeling, incorporating experimental data directly into the numerical simulations of boundary-value problems.
\smallskip\\
The data-driven scheme bypasses the empirical material modeling step by computing the closest point in the material data set consistent with the problem's compatibility and equilibrium condition. Consequently, it provides an alternative formulation of the classical initial-boundary-value problem based on nearest-neighbor clustering. 
\\
The approach has been fine-tuned for diverse applications: from non-linear elasticity \cite{kirchdoerfer:2016,kirchdoerfer:2017,conti:2018,nguyen:2018,galetzka:2020} to dynamics \cite{kirchdoerfer:2018} and finite strain \cite{platzer:2021}. It's also been adapted for material data identification \cite{stainier:2019}, non-local mechanics \cite{karapiperis:2019}, electro-mechanical problems \cite{marenic:2022}, homogenization schemes \cite{zschocke:2022}, and model-driven coupling \cite{yang:2022}. Ibañez et al. \cite{ibanez:2017,ibanez:2018} refined the approach using a manifold learning method that maps data into a lower-dimensional space to use the locally linear embeddings. Eggersmann et al. \cite{eggersmann:2021} presented a second-order data-driven approach that uses tensor voting \cite{mordohai:2010} to obtain point-wise tangent space, enabling the search for additional states close to the original data. 
For inelastic boundary value problems, Eggersmann et al. \cite{eggersmann:2019} include local histories in the data set to investigate materials with memory. Karapiperis et al. \cite{karapiperis:2021} have also suggested a variation of the scheme, considering multiscale modeling. In addition, we recently developed a paradigm incorporating the tangent space into the distance-minimizing data-driven formulation and classifies the underlying data structure into subsets according to various material behaviors \cite{ciftci:2022}. The framework features a parametrization of the material history and an optimal sampling of the mechanical system's state space.
\smallskip\\
The paradigm's dependence on the nearest-neighbor clustering of data points proposes research areas in machine-learning methods, particularly Artificial Neural Networks (ANNs), that are known to approximate any continuous function for appropriate network parameters \cite{csaji:2001,lu:2017}. The flexibility and quality of neural networks led to success in a wide range of applications, e.g., image recognition \cite{he:2016},  language processing \cite{lafferty:2001}, or generative modeling \cite{goodfellow:2014,goodfellow:2020}. An extension to neural networks is physics-informed deep learning, successfully used in solving physical-related problems such as fluid mechanics \cite{raissi:2020,sun:2020}, aerodynamics \cite{mao:2020,dourado:2020}, shell structures \cite{bastek:2023} or material science \cite{he:2020,yin:2021}. Physics-Informed Neural Networks (PINNs) can be trained to fulfill training data and learn optimal solutions for allocated physics-governing equations by specifying appropriate loss functions \cite{lagaris:1998,raissi:2019}. The physics-based loss competes against a data-based loss, which is needed to provide fundamental knowledge of the system. Thus, partial differential equations act as additional constraints during network training, resulting in a multi-objective optimization problem. Optimizing data and physics give physics-informed neural networks flexibility in solving forward and inverse problems \cite{dourado:2020,sun:2020,mao:2020,haghighat:2021,yang:2021,kadeethum:2020}. The trade-off between the individual losses can be influenced using hyper-parameter \cite{platt:1987,hwang2012,wang:2021}. For example, adaptive activation functions \cite{jagtap:2020,jagtap2020_2}, or manually weighted losses \cite{jin:2021}, can improve the quality of the neural network for specific problems. Another approach to overcome the local convergence issue due to global approximation is the usage of adaptive training strategies and domain decomposition \cite{henkes:2022}.
This investigation combines the model-free data-driven approach with Generative Adversarial networks (GANs). In machine learning, GANs have emerged as a powerful tool consisting of two neural networks – the generator, which creates data, and the discriminator, which evaluates the authenticity of the generated data. Through their adversarial game, GANs are adept at generating high-fidelity data, often indistinguishable from actual data \cite{goodfellow:2020}. 
An extension is the integration of physics-informed neural networks with the GAN structure. For instance, the pursuit of robust uncertainty quantification within the framework of PINNs has led to recent methodologies. The PIG-GAN framework \cite{yang:2019} harnesses the capabilities of a physics-informed generator to address adversarial uncertainty. On the other hand, the PID-GAN approach \cite{daw:2021} uses a physics-informed discriminator, carving out a distinct avenue to achieve reliable uncertainty quantification while maintaining fidelity to the governing physics. Another stride in this direction is the DeqGAN, which offers a unique perspective on PINNs by learning the loss function via generative adversarial networks. This methodology provides a robust avenue for solving the challenges traditionally associated with defining appropriate loss functions for PINNs \cite{bullwinkel:2022}.
In our approach, the generator is a  physics-informed neural network, and the discriminator employs the closest strain-stress data to evaluate the authenticity of the generator's results. This synergized methodology matches model-free data-driven computational mechanics and deep learning principles, aspiring to more accurately simulate and predict intricate mechanical behaviors. 

Section \ref{sec:ddriven} provides a general setting by introducing the definitions and derivation of the distance-minimizing data-driven computing method based on \cite{conti:2018}. Section \ref{sec:GAN_ddriven} introduces the framework of artificial neural networks and generative adversarial networks. In addition, we propose using a physics-informed GAN to solve the distance-minimizing data-driven problem. Section \ref{sec:num_example} exhibits the performance of the proposed method using a numerical example involving a non-linear elastic in-plane boundary value problem. Finally, Section \ref{sec:conclusion} summarizes the results and suggests future research subjects. 

\section{Model-free Data-driven setting}\label{sec:ddriven}
The following will summarize the classical data-driven computational mechanics
method for the reader's convenience based on the definitions and formulations in \cite{conti:2018}. We consider an elastic body $\Omega \subset \mathbb{R}^d$ whose internal states are defined by displacement field $\fu :\Omega \to \mathbb{R}^{d}$ and the compatibility and equilibrium conditions
\begin{equation}\label{eq:constraint}
\begin{aligned}
	&\feps(\fx) - \nabla^{\textrm{sym}} \fu(\fx) = \bold 0,	&  \text{in } \Omega, \\  
	&\nabla \cdot \fsig(\fx) - \ff(\fx) = \bold 0,	&  \text{in } \Omega,
\end{aligned}
\end{equation}
and boundary conditions
\begin{equation}\label{eq:bc_constraint}
\begin{aligned}
	&\fu(\fx) = \fg(\fx), &  \text{on } \Gamma_D ,\\
	&\fsig(\fx) \cdot \fn(\fx) = \ft(\fx), & \text{on } \Gamma_N ,
\end{aligned}
\end{equation}
where $\feps:\Omega \to \mathbb{R}^{{d}\times {d}}_{  \textrm{sym}}$ is the strain field and $\fsig:\Omega \to \mathbb{R}^{{d}\times {d}}_{  \textrm{sym}}$ is the stress field. The boundary $\Gamma$ of the domain $\Omega$ is defined by the Dirichlet ($\Gamma_D$) and Neumann ($\Gamma_N$) with $\Gamma = \Gamma_D \cup \Gamma_N$ and $\Gamma_D \cap \Gamma_N = \emptyset$.  In addition, $\ff:\Omega \to \mathbb{R}^{d}$ is the body force, and $\fg, \ft, \fn: \Gamma \to \mathbb{R}^{d}$ define the boundary displacement, applied traction and outer normal, respectively. \\
We define $Z_\textrm{loc} \subset  \mathbb{R}^{{d}\times {d}}_{\textrm{sym}}\times  \mathbb{R}^{{d}\times {d}}_{  \textrm{sym}}$ as the local phase space consisting of pairs $\fz(\fx) = (\feps(\fx), \fsig(\fx))$ describing the local state of the system at material point $\fx$. The global phase space $Z$ is defined as the collection of the state functions, i.e.
\begin{align}
Z = \{\fz \,:\,\fz \in Z_\textrm{loc}\}.
\end{align}
The data-driven distance-minimization problem, introduced by \cite{kirchdoerfer:2016}, reads
\begin{align}\label{eq:argmin_problem}
\argmin_{\fz\in \mathcal{C}, \hat{\fz} \in \mathcal{D}} d(\fz,\hat{\fz}),
\end{align}
where $\mathcal{C} \subset Z$ denotes the constraint set defined by
\begin{equation}
\mathcal{C} :=\Big\{\fz\in Z:\eqref{eq:constraint}\; \text{and}\; \eqref{eq:bc_constraint}\Big\};
\end{equation}
containing all states fulfilling compatibility and equilibrium. The set $\mathcal{D} \subset Z$ consists of a finite number of experimental measurements achieved from small-scale simulations and is defined by
\begin{equation}\label{eq:dataset}
\mathcal{D} = \{\fz \in Z\,:\, \fz(\fx) \in \mathcal{D}_\textrm{loc}\}
\quad \text{with}  \quad
\mathcal{D}_\textrm{loc} =\{(\feps_i,\fsig_i)\}_{i=1}^{n_e}, 
\end{equation}
where $n_e \in \mathbb{N}$ is the number of local data points associated with the material point. The distance $d:Z \times Z \to \mathbb{R}$ is defined by 
\begin{align}\label{eq:distance}
d(\fz, \hat{\fz}):= \|\fz - \hat{\fz} \|,
\end{align}
metricized by the norm
\begin{equation}\label{eq:norm}
\|\fz\|^2 :=\int\limits_{\Omega}\left(\frac{1}{2}\fC \feps \cdot \feps + \frac{1}{2}\fC^{-1}\fsig\cdot\fsig\right) \bold d\fx ,
\end{equation}
where $\fC \in \mathbb{R}^{{d}\times {d}}_{\textrm{sym}}$ is a symmetric positive definite matrix typically being of the type of elastic stiffness. Thus, the data-driven paradigm aims to find the closest point $\fz$ in the constraint set $\mathcal{C}$ to $\hat{\fz}$ in the material data set $\mathcal{D}$. 
\\ \\ 
Challenges such as data availability, noise, inconsistency, and high dimensionality frequently arise in the data-driven paradigm. Traditional analytical and computational methods may need to be adjusted when addressing these issues. Consequently, the incorporation of machine learning, particularly methods like generative adversarial networks coupled with physics-informed generators, is considered. This integration is aimed at effectively handling the complexities of data-driven datasets, ensuring the outcomes remain consistent with domain-specific knowledge. The following sections will present a detailed discussion on the principles of artificial neural networks and physics-informed neural networks, illustrating the approach of physics-informed generative adversarial networks to solve the data-driven boundary value problem \eqref{eq:argmin_problem}.

\section{Generative adversarial networks with physics-informed generators for model-free data-driven problems}\label{sec:GAN_ddriven}
This section delves into the application of Generative Adversarial Networks (GANs) equipped with Physics-Informed Generators for addressing the model-free data-driven problem. A GAN involves a competitive dynamic between two neural networks, forming a zero-sum game: one network's success implies the other's setback. To harness GANs for resolving the data-driven boundary value problem depicted in \eqref{eq:argmin_problem}, Section \ref{subsec:pinn} initiates with a concise overview of Artificial Neural Networks (ANNs) and explains physics-informed neural networks (PINNs). Section \ref{subsec:GAN} lays out the foundational principles of GANs, and in Section \ref{subsec:GAN_PINN}, we pivot to the novel approach of leveraging GANs augmented with PINNs to solve the data-driven boundary value problem.

\subsection{Physics-informed neural networks}\label{subsec:pinn}
\begin{figure}
\centering
\includegraphics[scale=0.6]{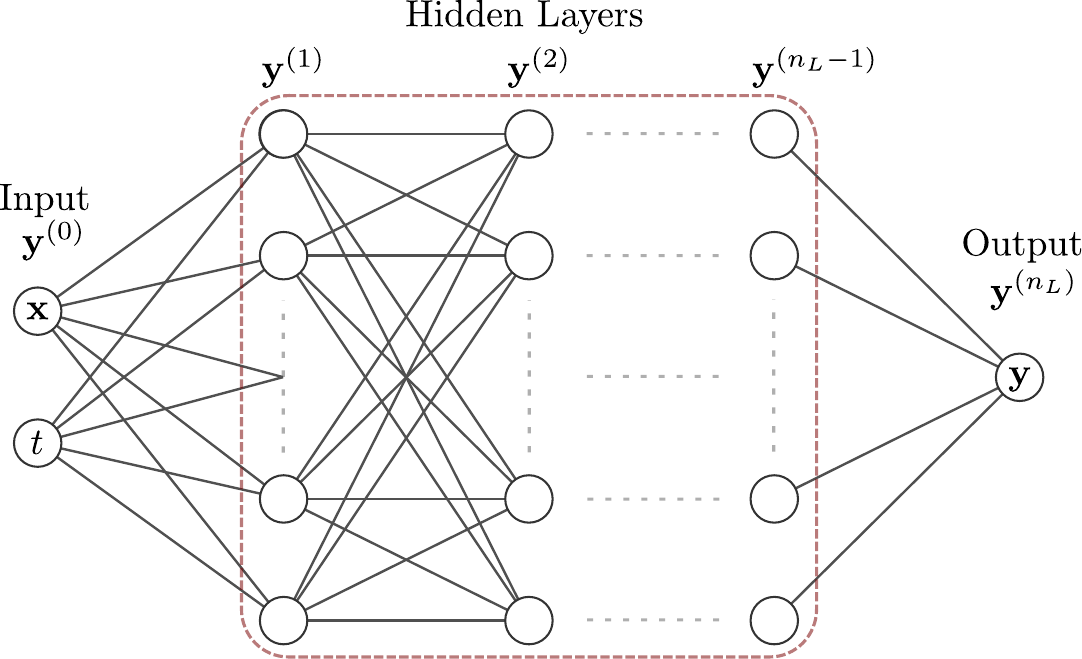}
\caption{Schematic representation of the neural network architecture $\mathcal{N}(\fx, t;\ftheta)$, starting with multi-input parameters $\fx$ and $t$ at input layer $\fy^{(0)}$, progressing through sequential hidden layer $\fy^{(1)}, \ldots, \fy^{(n_L -1)} $. The architecture concludes with an output layer $\fy^{(n_L)}$, producing the final output $\fy$.}
\label{fig:ann_topology}
\end{figure}

\begin{figure}[H]
\centering
\includegraphics[scale=0.57]{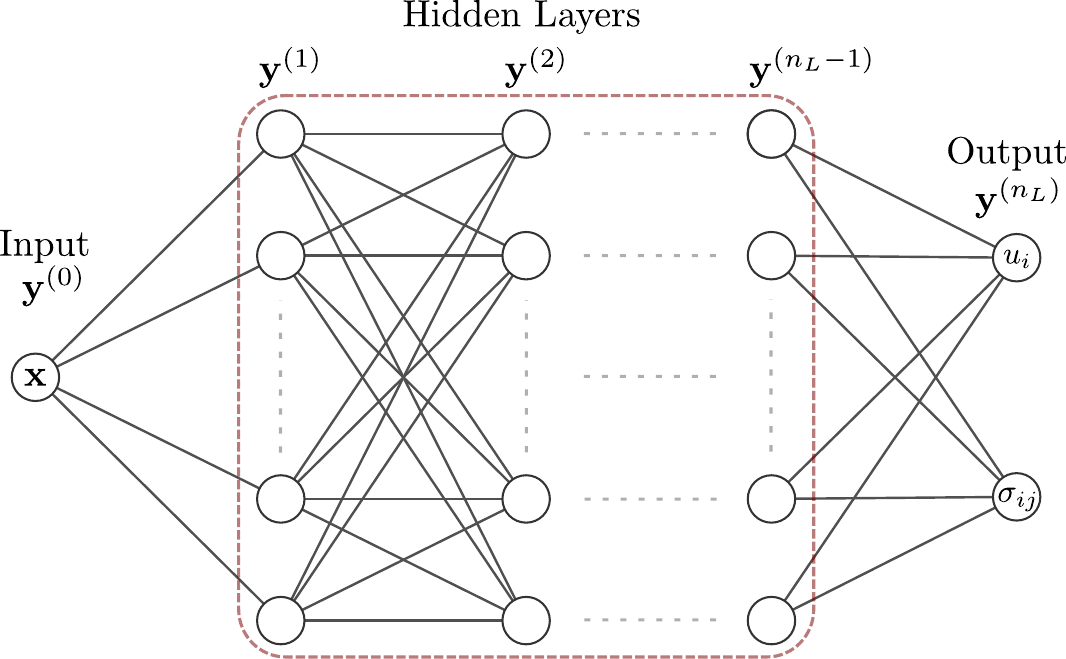}
\caption{Schematic representation of the neural network's topology, illustrating the progression from input through multiple hidden layers, resulting in displacement components $u_i$ and stress components $\sigma_{ij}$.}
\label{fig:ann_ds_topology}
\end{figure}

\begin{figure}[H]
\centering
\includegraphics[scale=0.3]{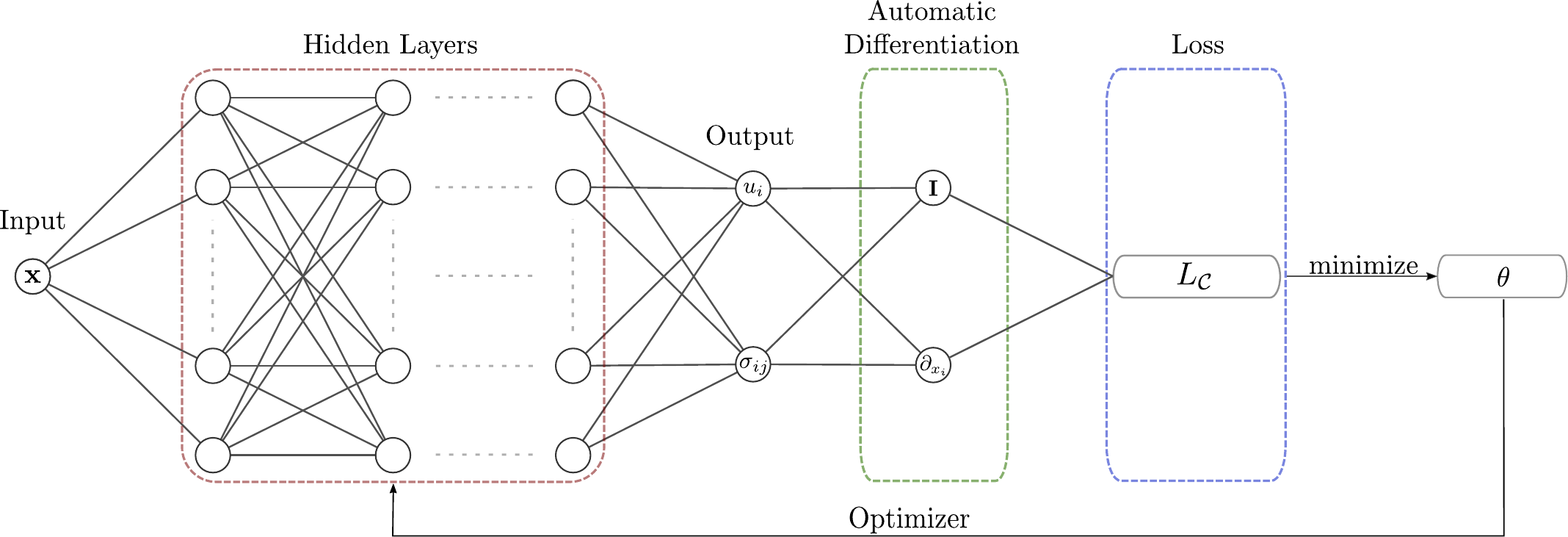}
\caption{Schematic representation of the physics-informed neural network's topology, illustrating the progression from input through multiple hidden layers, resulting in displacement components $u_i$ and stress components $\sigma_{ij}$. These outputs undergo automatic differentiation to compute a physics-based loss function $L_\mathcal{C}$ .
	The loss is minimized using an optimizer to refine the network parameters $\theta$.}
\label{fig:pinn_ds}
\end{figure}

Based on the universal function approximation theorem \cite{hornik:1989}, an artificial neural network is a parametrized, non-linear function composition that can approximate arbitrary Borel measurable functions. This section introduces the basic concept based on the definitions and formulations in \cite{henkes:2022}. For this purpose, we introduce a densely connected feed-forward neural network, denoted by the map $\mathcal{N}:\mathbb{R}^{d_x} \times [0, T] \to \mathbb{R}^{d_y}$, which is defined by a composition of  $n_L \in \mathbb{N}$ non-linear functions:
\begin{align}\label{eq:ANN}
\mathcal{N}:\mathbb{R}^{d_x} \times [0, T] &\to \mathbb{R}^{d_y}\\
(\fx, t) &\mapsto \mathcal{N}(\fx, t)=\fy^{(\ell)} \circ \ldots \circ \fy^{(0)} = \fy,
\end{align}
for $\ell = 1,\ldots, n_L$, where $\fx$ denotes the spatial part of the input vector of dimension $d_x \in \mathbb{N}$ at time $t \in [0, T]$ with $T>0$ and $\fy$ denotes the output vector of dimension $d_y \in \mathbb{N}$. In this context, $\fy^{(0)}$ and $\fy^{(n_L)}$ are called the input and output layer, such that
\begin{align}
\fy^{(0)} = (\fx, t), \qquad \fy^{(n_L)}  = \mathcal{N}(\fx, t).
\end{align}
The functions $\fy^{(\ell)}$ are called hidden layers and define a $\ell-$fold composition, mapping the input $(\fx, t)$ to the output $\fy$ by
\begin{align}
\fy^{(\ell)} = \{\fy^{(\ell)}_\eta,\,\eta=1,\ldots,\eta_u \}, \text{ with } \fy^{(\ell)}_\eta= \mathrm{act}^{(\ell)}\left(\bold W^{(\ell)}_\eta \fy^{(\ell-1)} + \bold b_\eta^{(\ell)}\right).
\end{align}
We call $\fy^{(\ell)}_\eta$ the $\eta^\text{th}$ neural unit of the $\ell^\text{th}$ layer $\fy^{(\ell)}$, where $\eta_u \in \mathbb{N}$ is the total number of neural units per layer.
$\bold W_\eta^{(\ell)}$ and $\bold b_\eta^{(\ell)}$ denote the weight matrix and bias vector of the $\eta^\text{th}$ neural unit in the $\ell^\text{th}$ layer $\fy^{(\ell)}$. Furthermore $\mathrm{act}^{(\ell)}(\cdot):\mathbb{R}\to \mathbb{R}$ is a non-linear activation function. 
All weights and biases of all layers $\fy^{(\ell)}$ are assembled in
\begin{align}
\ftheta= \Big\{\left(\bold W_\eta^{(\ell)}, \bold b_\eta^{(\ell)}\right);\, \ell = 1,\ldots,n_L,\, \eta= 1,\ldots, \eta_u \Big\},
\end{align}
including all parameters of the neural network $\mathcal{N}(\fx, t)$. As a result, the notation $\mathcal{N}(\fx, t;\ftheta)$ highlights the dependence of a neural network's output on the input and the current realization of the weights and biases. Figure \ref{fig:ann_topology} illustrates the network's topology, a combination of layers, neural units, and activation functions.
\\ \\
The main idea of solving boundary value problems with an artificial neural network is the reformulation to an optimization problem \cite{berg:2018,raissi:2019,cai:2022}, where the residual of the differential equations is to be minimized. To solve the differential equation \eqref{eq:constraint} and \eqref{eq:bc_constraint}, a suitable topology for the artificial neural network and, consequently, the physics-informed neural networks described in Section \ref{subsec:pinn} has to be chosen. Since \eqref{eq:constraint} is stationary, we can reduce the artificial neural network to  $\mathcal{N}(\fx;\ftheta)$. Thus, we can define neural networks as an ansatz for the displacement and stress field i.e.
\begin{align}
\fu(\fx, t) \approx \mathcal{N}_{u}(\fx; \ftheta_{u}), \\
\fsig(\fx, t) \approx \mathcal{N}_{\sigma}(\fx; \ftheta_{\sigma}),
\end{align}
with trainable network parameters $\ftheta:=\{\ftheta_u, \ftheta_\sigma\}$. Notably, there is no separate network for the strain tensor. The strain tensor is deduced using the kinematics and differentiation applied to the displacement network, i.e. $\feps = \nabla^{\textrm{sym}}\mathcal{N}_{u}(\fx; \ftheta_{u})$. The architecture of this artificial neural network is visualized in Fig.\ref{fig:ann_ds_topology}. 
Using the neural network ansatz we can rewrite the physics \eqref{eq:constraint} and \eqref{eq:bc_constraint} as
\begin{equation}
\begin{aligned}
	R_\Omega &= \nabla \cdot \mathcal{N}_\sigma(\fx, \ftheta_\sigma) - \ff(\fx),&& \text{in } \Omega, \\ 
	R_{\Gamma_D}  &= \mathcal{N}_u(\fx, \ftheta_u) - \fg(\fx), &&  \text{on } \Gamma_D ,\\
	R_{\Gamma_N} &= \mathcal{N}_\sigma(\fx, \ftheta_\sigma) \cdot \fn(\fx) - \ft(\fx), 	&&  \text{on } \Gamma_N,
\end{aligned}
\end{equation}
where $	R_\textrm{PDE}$ penalizes the residual of the equilibrium equation, and the equations $R_{\Gamma_D}$ and $R_{\Gamma_N}$ describe the discrepancy of the Dirichlet and Neumann boundary conditions. Notice that if $\mathcal{N}_u$ and $\mathcal{N}_\sigma$ is a solution to the original boundary value problem, they minimize the differential equation-based residuals.
\\
The parameters $\ftheta$ of the networks can be found by incorporating the physics-induced residuals into the training process of a neural network as components of the loss function. For this, we use a collocation method discretizing the domain $\Omega$ and the boundary $\Gamma:=\Gamma_D \cup \Gamma_N$ into sets of sample points $S_\Omega$ and $S_\Gamma$ with cardinalities $\vert S_\Omega \vert$ and $\vert S_\Gamma \vert$. Then, an optimization problem to find the optimal parameters $\ftheta^\star$, also called training, is defined as
\begin{align}\label{eq:loss_argmin}
\ftheta^\star &=\argmin\limits_{\ftheta} L_\mathcal{C} 
\end{align}
with $L_\mathcal{C}:=L_\Omega(\fx, \ftheta)  + L_{\Gamma}(\fx, \ftheta)$ given by the local losses
\begin{align}\label{eq:loss_phy}
L_{\Omega} &=  \frac{1}{\vert S_\Omega\vert} \sum_{\fx\in S_\Omega} \|R_\Omega(\fx; \ftheta)\|^2_2, \\
L_{\Gamma} &=  \frac{1}{\vert S_\Gamma\vert}\left(\sum_{\fx\in S_{\Gamma_D}} \|R_{\Gamma_D}(\fx; \ftheta)\|^2_2, + \sum_{\fx\in S_{\Gamma_N}} \|R_{\Gamma_N}(\fx; \ftheta)\|^2_2\right). 
\end{align}
The expressions penalize the residual of the governing equations and the discrepancy of the Dirichlet and Neumann boundary conditions, respectively. Notice that in the three-dimensional setting, one defines the neural networks as tuples, i.e.
\begin{align}\label{eq:nn_ds}
\mathcal{N}_u(\fx; \ftheta_u)  &= \{\mathcal{N}_{u_i}(\fx; \ftheta_{u_i}) \, \vert \, i = 1,2\}, \\
\mathcal{N}_\sigma(\fx; \ftheta_\sigma)  &= \{\mathcal{N}_{\sigma_{ij}}(\fx; \ftheta_{\sigma_{ij}})\, \vert \, i,j = 1,2,3 \text{ and } ij = ji\},
\end{align}
including the three components $u_i$ of displacement $\fu$ and six stress components $\sigma_{ij}$, where $ij=ji$ ensures the symmetry of the stress tensor $\fsig$. Fig. \ref{fig:pinn_ds} illustrates the complete network's structure. 
While the PINN framework provides a straightforward method to solve physical-enhanced problems, it has challenges. Notably, there have been instances where the optimization yields solutions with unexpected or non-physical behaviors even when carefully tailored to encapsulate the physics \cite{bastek:2023}. Additionally, the current PINN formulation must minimize the difference between the network's outputs and the available strain-stress data $\mathcal{D}$ due to the nature of the data-driven distance minimization problem \eqref{eq:argmin_problem}.
If we integrate the distance as an additional loss into the global loss, the whole problem becomes a nested optimization, leading to training challenges. The neural network could optimize in an undesired direction during each training epoch. If the approximated strain-stress point is not accurate, the corresponding data point might be suboptimal concerning the optimization algorithm, further complicating the learning process.
To address these challenges, we consider the integration of PINNs with generative adversarial networks. GANs are proficient at generating outputs with the same properties as actual data, providing a potential approach to generating realistic strain-stress solutions. Their flexibility ensures adaptability across diverse data types suited for various physical conditions. Moreover, the inherent capability of GANs to discern and capitalize on intricate patterns may lead to a more robust representation of underlying physics. Additionally, with conditional GANs, generating outputs based on specific conditions becomes feasible, allowing for more targeted solutions. The combined PINN-GAN approach seeks to ensure physical consistency and alignment with observed data, leveraging the strengths of both methodologies. For clarity, we will provide a brief overview of GAN theory in the following.

\subsection{Intermezzo to generative adversarial networks}\label{subsec:GAN}
\begin{figure}[H]
\centering
\includegraphics[scale=0.1]{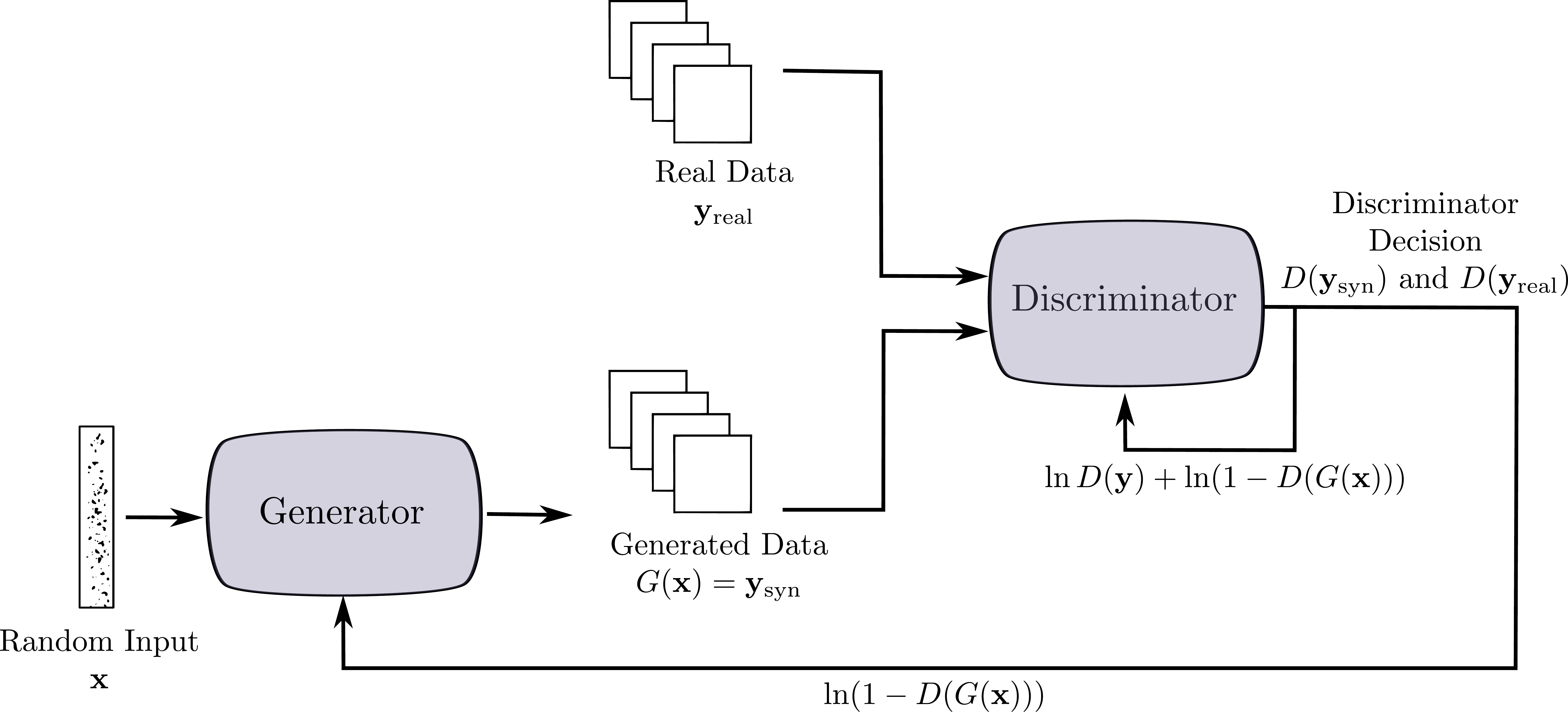}
\caption{
	Schematic representation of a generative adversarial network (GAN) showcasing the interaction between the generator producing data from random input and the discriminator evaluating the authenticity of both real and generated data.
}
\label{fig:GAN_architecture}
\end{figure}

Introduced by Goodfellow et al.~\cite{goodfellow:2014}, generative adversarial networks illustrate a novel approach to generating data using neural architectures. These networks comprise two distinctive neural entities: the generator ($G$) and the discriminator ($D$). The underlying goal of a GAN is to generate data instances that emulate the properties of actual data. The generation is achieved by setting the two networks against each other in a competitive game, often described as a dual-player minimax game.
\\\\
Taking reference from the definitions provided in \eqref{eq:ANN}, we define the real data space as $\mathbb{D}_{\text{real}} \subset \mathbb{R}^{d_y},$ where $d_y$ is the dimension of the space, i.e., $d_y = \dim(\mathbb{D}_{\text{real}}).$ The main objective of GANs is to produce synthetic data denoted as $\mathbf{y}_{\mathrm{syn}}$, residing in the same space as our real data $\mathbf{y}_{\text{real}}$. The generator can be defined as a function $G:\mathbb{R}^{d_x} \to \mathbb{R}^{d_y},$ which transforms a random noise vector $\mathbf{x}$ into synthetic data $\mathbf{y}_{\mathrm{syn}}$. In contrast, the discriminator operates as a function $D:\mathbb{R}^{d_y} \to \mathbb{R},$ that provides a measure of authenticity for a given data sample. Mathematically, these networks can be illustrated as:
\begin{align}
\begin{split}
	G:\mathbb{R}^{d_x} &\to \mathbb{R}^{d_y} \\
	\mathbf{x} &\mapsto \mathcal{N}_G(\mathbf{x}; \boldsymbol{\theta}_G),
\end{split} 
\qquad \text{and} \qquad 	
\begin{split}
	D:\mathbb{R}^{d_y} &\to [0, 1]  \\
	\mathbf{y} &\mapsto  \mathcal{N}_D(\mathbf{y}; \boldsymbol{\theta}_D).
\end{split}
\end{align}
Here, $\mathcal{N}_G(\mathbf{x}; \boldsymbol{\theta}_G)$ and $ \mathcal{N}_D(\mathbf{y}; \boldsymbol{\theta}_D)$ describe the neural networks with their corresponding trainable parameters $\boldsymbol{\theta}_G$ and $\boldsymbol{\theta}_D$.
The adversarial game between the generator and the discriminator during training can be encapsulated in the following objective
\begin{align}\label{eq:GAN_loss}
L(G, D) = \mathbb{E}_{\mathbf{y} \sim p_{\text{data}}}[\ln D(\mathbf{y})] + \mathbb{E}_{\mathbf{x} \sim p_{\mathbf{x}}}[\ln(1 - D(G(\mathbf{x})))],
\end{align}
leading to the optimization:
\begin{align}\label{eq:minmax_L}
\min_{G} \max_{D} L(G, D),
\end{align}
where \(\mathbb{E}\) represents a random variable's expectation or expected value. It provides a weighted average of a function concerning its probability distribution. Specifically, 
\begin{align}
\mathbb{E}_{\mathbf{y} \sim p_{\text{data}}}[\ln D(\mathbf{y})] 
\end{align}
represents the average logarithmic score assigned by the discriminator to actual data samples drawn from the distribution \(p_{\text{data}}\). On the other hand, the expression 
\begin{align}
\mathbb{E}_{\mathbf{x} \sim p_{\mathbf{x}}}[\ln(1 - D(G(\mathbf{x})))]
\end{align} 
reflects the average logarithmic score the discriminator accords to the synthetic or generated data, which is created from a random noise vector \(\mathbf{x}\) following the noise distribution \(p_{\mathbf{x}}\).
\\ 
The competition between the two networks is straightforward: the generator $G$ aims to produce data that the discriminator $D$ cannot distinguish from accurate data. In contrast, the discriminator tries to better distinguish real data from fake data produced by $G$. The probability distributions $p_{\text{data}}$ and $p_{\mathbf{x}}$ depict the actual data and noise distributions, respectively. The terms in the objective function essentially capture the average confidence levels of the discriminator in judging the authenticity of both original and fake data samples. The procedure of the GAN's interplay between the generator and the discriminator is illustrated in Fig.~\ref{fig:GAN_architecture}.
Traditional GANs deploy a sigmoid activation function for the discriminator's final layer, ensuring its outputs fall within [0,1]. The GANs can suffer from issues like mode collapse (where the generator generates limited varieties of samples), vanishing gradients, and general training instability. To address some of these challenges, the Wasserstein GAN (WGAN) \cite{arjovsky:2017} changes the objective function to leverage the Wasserstein distance \cite{kantorovich1960mathematical}. The WGAN objective can be described as:
\begin{align}\label{eq:WGAN_L}
L_{WGAN}(G, D) = \mathbb{E}_{\mathbf{y} \sim p_{\text{data}}}[D(\mathbf{y})] - \mathbb{E}_{\mathbf{x} \sim p_{\mathbf{x}}}[D(G(\mathbf{x}))],
\end{align}
leading to the following optimization:
\begin{align}\label{eq:minmax_L_WGAN}
\min_{G} \max_{D} L_{WGAN}(G, D).
\end{align}
WGANs are known to provide more stable and consistent training dynamics~\cite{arjovsky:2017}. Building on the WGAN, the Wasserstein GAN with Gradient Penalty (WGAN-GP) introduced a regularization term to ensure that the discriminator's gradients remain bounded \cite{mescheder2018training}. This gradient penalty aims to enforce the Lipschitz continuity condition, which addresses the vanishing gradient problem. The gradient penalty is defined as:
\begin{align}
\text{GP} = \mathbb{E}\left[ \left( \lVert \nabla_{\tilde{\mathbf{y}}} D(\tilde{\mathbf{y}}) \rVert_2 - 1 \right)^2 \right],
\end{align}                                                          
where $\tilde{\mathbf{y}} = \delta \mathbf{y}_{\text{real}} + (1 - \delta) \mathbf{y}_{\mathrm{syn}}$ and $\delta$ is sampled from a uniform distribution in $[0,1]$. The optimization for WGAN-GP thus becomes
\begin{align}
\min_{G} \max_{D} L_{WGAN}(G, D) + \omega  \cdot\text{GP},
\end{align}
where $\omega \in \mathbb{R}_+$ is a hyperparameter determining the weight of the gradient penalty in the overall objective~\cite{gulrajani:2017}. 

\subsection{Physics-informed GANs for data-driven mechanics problems}\label{subsec:GAN_PINN}
In the classical data-driven computational mechanics paradigm Section \ref{sec:ddriven}, the objective is to find the closest point $\fz$ in the constraint set $\mathcal{C}$ to $\hat{\fz}$ in the material dataset $\mathcal{D}$, as formalized in equation \eqref{eq:argmin_problem}. This context motivates our modified GAN approach for data-driven mechanics problems. To utilize GANs for solving differential equations in a data-driven mechanics setting, we propose a novel approach wherein the generator in the GAN architecture is identified as a physics-informed neural network (PINN). In this paradigm, while the generator outputs plausible solutions adhering to the underlying physics, the discriminator is trained to distinguish between the generator's predictions and actual strain-stress data. 
\\
In the conventional GAN setup from Section \ref{subsec:GAN}, the generator $G$  maps the input vector $\mathbf{x}$ into synthetic data, $\mathbf{y}_{\mathrm{syn}}$. Instead of treating $\mathbf{x}$ as a random noise vector, it represents the collocation points $\fx$ in the domain $S_\Omega$. Thus, the generator is formalized as a mapping $G: S_\Omega \to (\mathcal{N}_u, \mathcal{N}_\sigma)$, where $\mathcal{N}_u$ and $\mathcal{N}_\sigma$ represents the neural network approximation for the displacement and stress field, respectively. Therefore, the generator can be defined as:
\begin{align}
G(\fx, \ftheta_G):= (\mathcal{N}_u(\fx; \ftheta_u),\mathcal{N}_\sigma(\fx; \ftheta_\sigma))
\end{align}
where $\ftheta_G := (\ftheta_u, \ftheta_\sigma)$ denotes the trainable parameters of the generator network. Building upon the physics-informed aspect, we differentiate $\mathcal{N}_u$ and employ the kinematics equation to obtain the strain $\feps$. Given $\feps = \nabla^\textrm{sym}\mathcal{N}_u$, the generator's output evolves from merely the neural network predictions  $\mathcal{N}_u$ and $\mathcal{N}_\sigma$ to the strain-stress pair $\fz :=(\feps, \mathcal{N}_\sigma)$. 

\begin{figure}[H]
\centering
\includegraphics[scale=0.1]{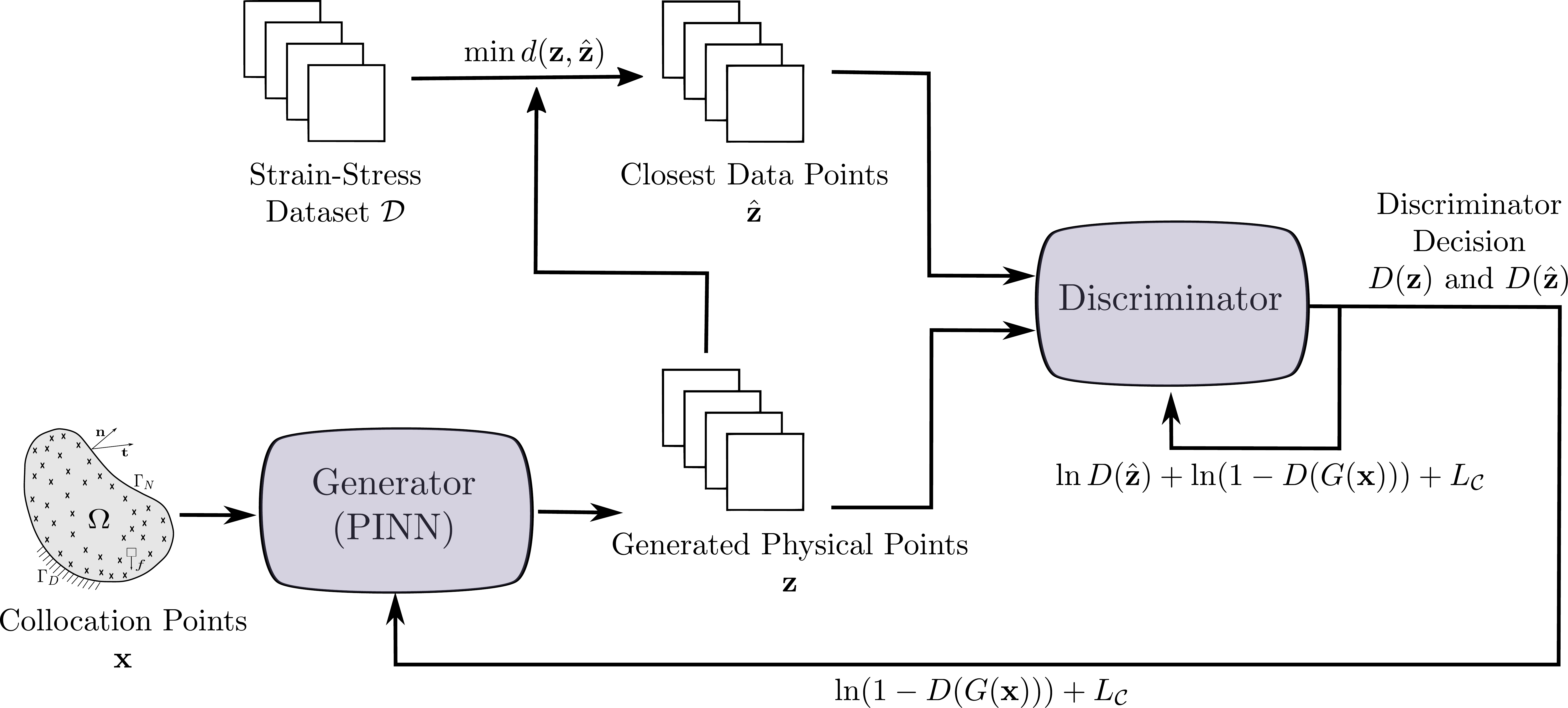}
\caption{ Schematic representation of a physics-informed generative adversarial network (GAN) incorporating collocation points and strain-stress data for physical point generation and discrimination.
}
\label{fig:GAN_PINN}
\end{figure}

Once we obtain the strain-stress output from the generator, to stay consistent with the data-driven mechanics' paradigm, we compute the strain-stress data points $\hat{\fz} \in \mathcal{D}$  closest to the output $\fz$, which corresponds to:
\begin{equation}
\hat{\fz} =\argmin_{\hat{\fz} \in \mathcal{D}} d(\fz,\hat{\fz}),
\end{equation}
with distance \eqref{eq:distance}. We then use $\fz$ and $\hat{\fz}$ as synthetic and real data for the discriminator's training. For the discriminator $D(\fy, \ftheta_D)$, we establish the mapping $D: \mathbb{R}^{2d} \to [0,1]$, aligning with the conventional GAN framework. To accommodate strain-stress pairs as inputs for the discriminator, we convert a pair into a $2d$-vector $\fy$ by applying Voigt-Notation to both the strain and stress, then merging them into a single vector. Given strain-stress data $\hat{\fz}\in \mathcal{D}$, it assesses the data's authenticity, furnishing scores to guide the generator's training. With the generator now representing a PINN, the adversarial loss in equation \eqref{eq:GAN_loss} has to integrate the physics-informed loss $L_\mathcal{C}$, derived from the residuals of the governing differential equations:
\begin{equation}
L(G, D) = \mathbb{E}_{\hat{\fz} \sim p_{\mathcal{D}}}[\ln D(\hat{\fz})] + \mathbb{E}_{\fx \sim p_{S_\Omega}}[\ln(1 - D(G(\fx))) +  L_\mathcal{C}].
\end{equation}
The collaborative training between the discriminator and the physics-informed generator ensures that the latter learns to craft data that confounds the discriminator and aligns closely with intrinsic physics. Fig.~\ref{fig:GAN_PINN} illustrated the physics-enhanced GAN approach for the data-driven mechanics problem. Regarding Wasserstein GANs and their gradient penalty variants, their objectives concerning the physics-informed generator must be modified. For instance, with the Wasserstein GAN objective, the loss function becomes:
\begin{equation}
L_\textrm{WGAN}(G, D) = \mathbb{E}_{\hat{\fz} \sim p_{\mathcal{D}}}[D(\hat{\fz})] - \mathbb{E}_{\fx \sim p_{S_\Omega}}[D(G(\fx)) + L_\mathcal{C}],
\end{equation}
Moreover, for the WGAN-GP, the combined objective is:
\begin{equation}
L(G, D) = L_\textrm{WGAN} + \omega \cdot \text{GP}.
\end{equation}
By incorporating GANs with physics-informed principles, the models produce data that adheres to the statistics of observed datasets and the underlying differential equations. This integration addresses the nested optimization issue commonly found in the PINN-based data-driven mechanics. With the capability of GANs to generate outputs mirroring accurate data, the solutions are both statistically relevant and in line with physical principles. Using GANs simplifies the optimization process, making the training more stable and less prone to errors from inaccurate strain-stress approximations. However, it is worth noting that the loss values obtained while training a traditional GAN are often unreliable. In many studies, qualitative and quantitative evaluation methods are employed to assess the performance of the GAN. Qualitative evaluations, while offering a quick visual validation, can be subjective. Typically, they involve human observers who evaluate the realism of a generated sample. The overall presumption has been that if the generated sample appears realistic, the GAN's training is deemed successful, regardless of potential fluctuations in loss values. Nevertheless, such evaluations can be biased and do not always represent the complete performance spectrum of the GAN. For instance, the generated samples might still appear high quality even in scenarios where mode collapse occurs.
Considering these challenges, especially in the context of our work where the goal is not generating images but accurately representing strain-stress states, we decided on WGAN + GP. Unlike traditional GANs, the loss of WGANs has a convergence point. Ideally, this point is reached when the generator is so adept at producing samples that no Lipschitz continuous discriminator can differentiate between real and generated samples. This characteristic of WGAN provides a more stable and consistent evaluation metric, ensuring that the generated strain-stress states are physically accurate. The effectiveness of this method will be showcased in a two-dimensional numerical example.

\section{Numerical benchmark of a non-Linear elastic plate with  hole}\label{sec:num_example}
This section illustrates the application of GANs to the data-driven computing paradigm \cite{kirchdoerfer:2016} in a typical benchmark, considering stress analysis of non-linear elastic material. We discuss the problem setup and test environments and give a proper definition of the geometry and boundary conditions and the material parameters for data generation. We limit the simulation to noiseless synthetic data sets, which consist of strain-stress points created numerically using a material model rather than obtained by actual experimental measurements. However, experimental data is often noisy and contains outliers. This issue can be addressed with noise reduction algorithms such as tensor voting \cite{kim:2013}, Kalman filtering \cite{kalman:1960}, and deep learning-based techniques.
\\ \\
In this benchmark, we investigate a $2d$ in-plain plate with a hole subject to a distributed force. The geometry, boundary conditions, and displacements are chosen according to a similar test presented in \cite{eggersmann:2021} and illustrated in Fig. \ref{fig:nl_plate_geometry}. 
\subsubsection*{Geometry:}\label{subsec:nl_plate_geometry}
\begin{figure}[H]
\centering
\includegraphics[scale=0.17]{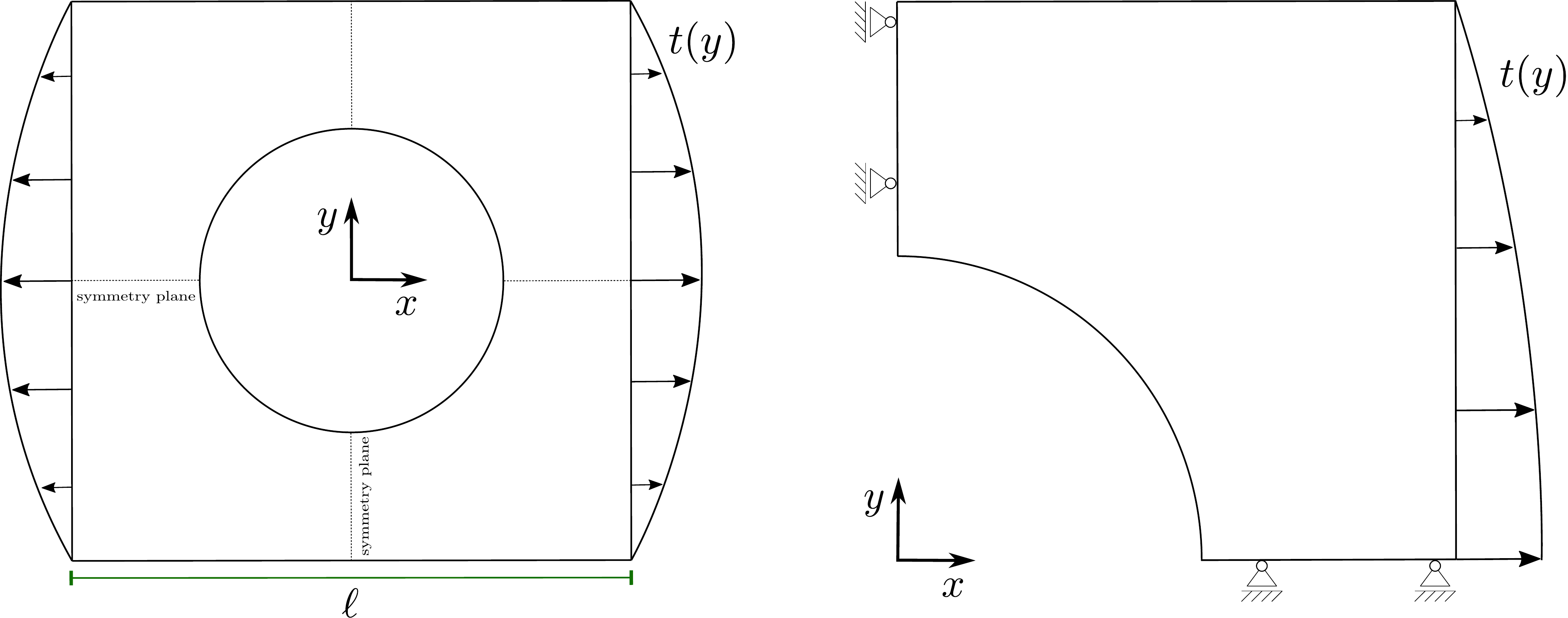}
\caption{Illustration of a square plate subjected to external forces, alongside its top-right quadrant representing the symmetry section with specified boundary conditions and force distribution $t(y)$ applied.}
\label{fig:nl_plate_geometry}
\end{figure}

The system is defined by $\Omega = \big[-\frac{\ell}{2}, \frac{\ell}{2}\big]^2 \setminus B_{r}(0)$, where $B_{r}$ refers to the open ball of radius $r = \frac{\ell}{4}$ centered at the origin $(0,0)$. The side lengths of the plate are equal to $\ell=2\textrm{m}$. Due to the symmetry of geometry, only one-quarter of the system is simulated, cf. Fig~\ref{fig:nl_plate_geometry}. 
Displacements are fixed at the quarter plate's left surface $x = 0$ in $x$-direction and at the bottom surface $y = 0$ in $y$-direction. The corresponding conditions read as follows:
\begin{equation}
\begin{aligned}	\label{eq:plate_bc1}
	\begin{cases}
		u_x =  0, & \text{if } x = 0;\\
		u_y = 0 , & \text{if } y = 0;\\
	\end{cases}
\end{aligned}
\end{equation}
where  $\sigma_x$ is the stress and $u_x$ and $u_y$ are the displacements in $x$ and $y$-directions, respectively. In addition, we define boundary conditions for the stress, especially for $x=\frac{\ell}{2}$ the plate is subjected to a distributed force $t(y) = 200\cos(\frac{\pi y}{2})$ in $x$-direction. The boundary conditions for the stress components read
\begin{equation}
\begin{aligned}	\label{eq:plate_bc2}
	\begin{cases}
		\sigma_{xx} = t(y), & \text{if } x = 1; \\
		\sigma_{yy} = 0 , & \text{if } y = 1;\\
		\sigma_{xy} = 0, & \text{if } (x,y) \in \partial\Omega.
	\end{cases}
\end{aligned}
\end{equation}
Notice that numerical methods based on the weak form of a boundary value problem innately satisfy shear-free boundary conditions on free boundaries. However, our PINN approach utilizes the strong form of the boundary value problem, so it is crucial to impose the zero stress boundary conditions directly \cite{henkes:2022}. 
\\ \\
To train the network, we utilize $128^2$ quasi-random points produced using the Sobol sequence \cite{sobol:1967}. For testing, we generate $256^2$ domain points using a uniform random distribution.

\subsubsection*{Material parameters:}\label{subsubsec:nl_plate_matmodel}
The boundary value problem considers the non-linear elastic material behavior of \cite{eggersmann:2021} defined by
\begin{align}\label{eq:nl_model}
\fsig = \lambda g(\mathrm{tr}(\feps))\boldsymbol{I} + \mu\feps + \fC\feps,
\end{align}
with $g(x)=((\vert x\vert + a)^p - a^p)\mathrm{sgn}(x)$ and $a,p \in \mathbb{R}$.
The applied material parameters are Young's modulus $E$, Poisson's ratio $\nu$, and orthotropic elasticity tensor for plane strain given by 
\begin{align}\label{eq:block_elas_matrix}
\fC= \begin{pmatrix}
	C_{11} && 2\nu (\bar{\lambda}+G_\perp) && 0 \\
	2\nu (\bar{\lambda}+G_{\perp}) && \bar{\lambda} + 2G_\perp && 0 \\
	0 && 0 && G_\parallel
\end{pmatrix},
\end{align}
where $\lambda = \frac{E\nu}{(1+\nu)(1-2\nu)}$ and $\mu = \frac{E}{2(1+\nu)}$ are the well known Lamé constants and \(C_{11} = 4.6875E\), \(G_{\perp} = 0.3E\), \(G_{\parallel} = 0.2E\) and \(\bar{\lambda} = \frac{2\nu^2+1}{15-20\nu^2}E\) are additional material parameters. The exact parameter values used for the reference solution and synthetic data are given in Table \ref{table:nl_plate_mat_parameter}.
\begin{table}[ht]
\centering
\begin{tabular}[t]{lcccc}
	\hline
	&$E \,[\text{MPa}]$  & $\nu\,[-] $ &$a\,[-]$&$p\,[-]$ \\
	\hline 
	& \\[\dimexpr-\normalbaselineskip+2pt]
	&$1 \times  10^4$& $0.3$  &$0.001$ & $0.005$ \\
	\hline
\end{tabular}
\caption{Material parameters}\label{table:nl_plate_mat_parameter}
\end{table}%

\subsubsection*{Synthetic data:}\label{subsubsec:nl_plate_data}
In order to simulate actual experimental measurements, we generate data artificially using the non-linear material model \eqref{eq:nl_model} based on the given material parameters. We investigate normal data distributions of $100^3$ strain-stress data points with a fixed random seed. The data is created by a zero-mean normal distribution with a standard deviation of $0.005$ in all strain dimensions.
\subsubsection*{WGAN parameter:}\label{subsubsec:nl_plate_nn}
For the adversarial network, the model consists of the generator and the discriminator setup.
The generator 
\begin{align}
G(\fx, \ftheta_G) = \{\mathcal{N}_{u}(\fx, \ftheta_{u}), \mathcal{N}_{\sigma}(\fx, \ftheta_{\sigma})\}
\end{align}
with $\fx = (x,y)$, $\ftheta_G =(\ftheta_u, \ftheta_\sigma)$ and
\begin{align}
\mathcal{N}_{u}(\fx, \ftheta_{u}) &= \{\mathcal{N}_{u_i}(\fx, \theta_{u_i})\, \vert \, i = x,y\}, \\
\mathcal{N}_{\sigma}(\fx, \ftheta_{\sigma}) &= \{\mathcal{N}_{\sigma_{ij}}(\fx, \theta_{\sigma_{ij}})\, \vert \, i,j = x,y\} ,
\end{align}
is constructed with a series of fully connected layers. The architecture utilizes $4$ hidden layers, each with $64$ neurons. The activation function used across these layers is the Swish function, defined as 
\begin{align}
\textrm{Hardswish}(x) = \begin{cases}
	0 & \text{if } x \leq -3, \\
	x & \text{if } x \geq 3,\\
	\frac{x^2+3x}{6} & \text{otherwise}.
\end{cases}
\end{align}
In addition, to optimize the network training, we hard enforce the boundary conditions from \eqref{eq:plate_bc1} and \eqref{eq:plate_bc2}, such that the output of the generator is given by

\begin{align}
\begin{aligned}
	\mathcal{N}_{u_x}(\fx; \theta_{u_x}) &= x \cdot \hat{\mathcal{N}}_{u_x}(\fx;\theta_{u_x}), \\
	\mathcal{N}_{u_y}(\fx; \theta_{u_y}) &= y \cdot \hat{\mathcal{N}}_{u_x}(\fx; \theta_{u_y}), \\
	\mathcal{N}_{\sigma_{xx}}(\fx; \theta_{\sigma_{xx}}) &= (1-x) \cdot \hat{\mathcal{N}}_{\sigma_{xx}}(\fx; \theta_{\sigma_{xx}}), \\
	\mathcal{N}_{\sigma_{yy}}(\fx; \theta_{\sigma_{yy}}) &= (1-y) \cdot \hat{\mathcal{N}}_{\sigma_{yy}}(\fx; \theta_{\sigma_{yy}}), \\
	\mathcal{N}_{\sigma_{xy}}(\fx; \theta_{\sigma_{xy}}) &= xy(x^2 + y^2 -0.25)\cdot \hat{\mathcal{N}}_{\sigma_{xy}}(\fx; \theta_{\sigma_{xy}}),
\end{aligned}
\end{align}
with $\fx = (x,y)$ and $\ftheta = (\theta_{u_i}, \theta_{\sigma_{ij}})$ being the tuple of all trainable network parameters regarding the displacement and stress component. In order to obtain the strains and optimize the loss function, the spatial derivatives are obtained by automatic differentiation. 
\\
On the other hand, the discriminator $D(\fy, \ftheta_D)$ comprises a network architecture of $3$ hidden layers, each with $16$ neurons, which uses the LeakyReLU activation function defined as
\begin{align}
\textrm{LeakyReLU}(x) = \begin{cases}
	x & \text{if } x \geq 0, \\
	\alpha x & \text{if } x < 0,\\
\end{cases}
\end{align}
with a slope of $\alpha = 0.2$ for negative values. 
%This implementation employs the conventional GAN and the WGAN + GP approaches. The distinguishing feature of the conventional GAN is the application of a sigmoid activation function at the output layer of the discriminator, defined as:
%\begin{align}
%	\textrm{Sigmoid}(x) = \frac{1}{1+e^{-x}}.
%\end{align} 
Regarding optimization, both the generator and the discriminator use the ADAM optimizer with a learning rate of $0.02$. The beta values for the moment estimates are set as $(0.5, 0.999)$. A learning rate scheduler is employed with a maximum learning rate of $0.02$. It is set to adjust the rate over a total of $200$ steps for both the Generator and Discriminator. 

\subsubsection*{Result:}\label{subsubsec:nl_plate_result}
\begin{figure}[H]
\centering
\includegraphics[scale=0.28]{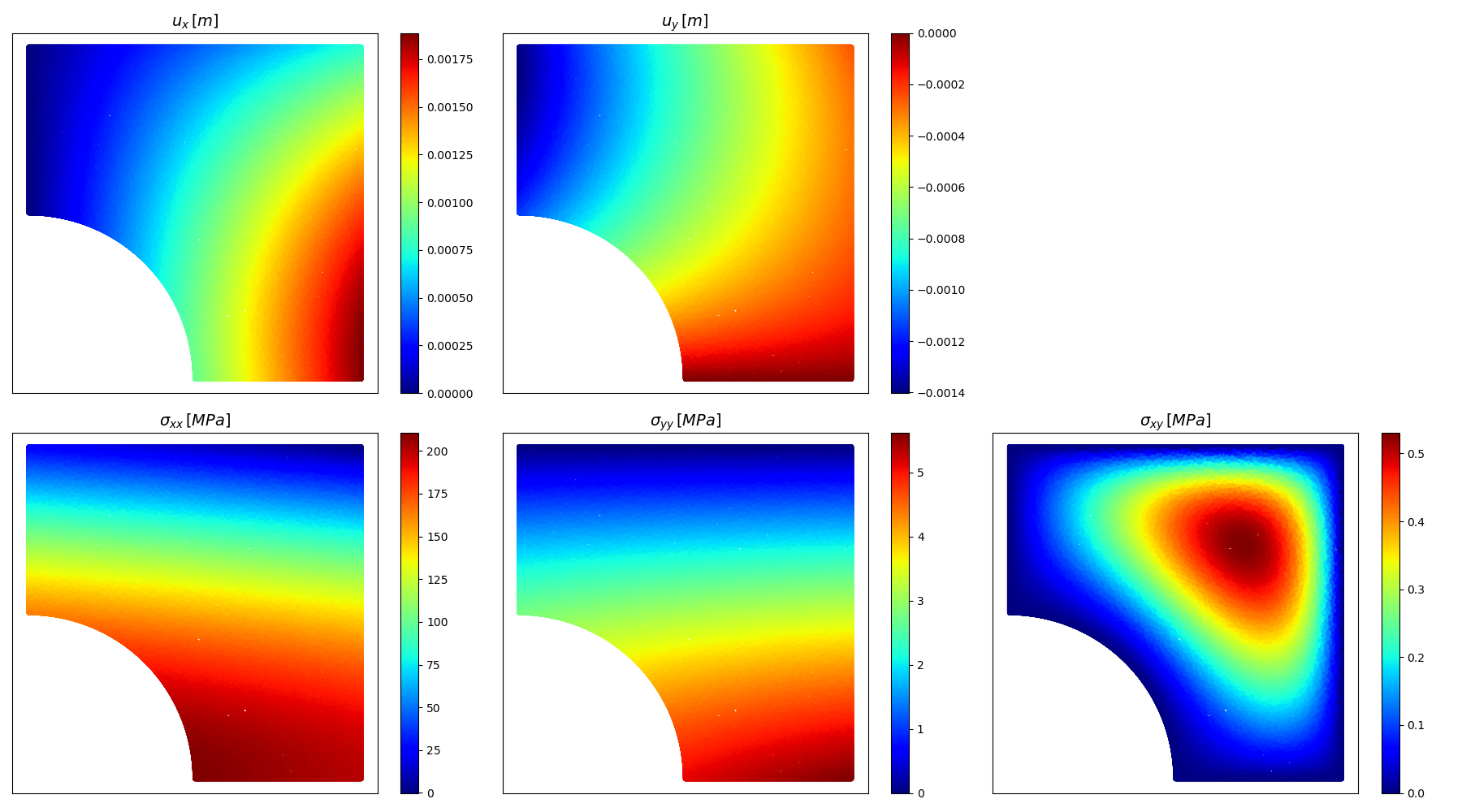}
\caption{Visualization of displacement and stress distribution after $200$ training epochs offering insights into the material's behavior under the applied loads and conditions \eqref{eq:plate_bc1} and \eqref{eq:plate_bc2}. From top-left to bottom-right: $u_x$ showcases a gradient, indicating a maximum displacement at $(x,y)=(1,0)$; $u_y$ reveals a displacement trend with negative values highlighted in $x=0$;  $\sigma_x$ shows a maximum stress magnitude at $y=0$; $\sigma_y$ displays a similar gradient; and $\sigma_{xy}$ captures a pronounced shear stress distribution inside the plate.}
\label{fig:wgan_strain_stress}
\end{figure}

\begin{figure}[H]
\centering
\includegraphics[scale=0.8]{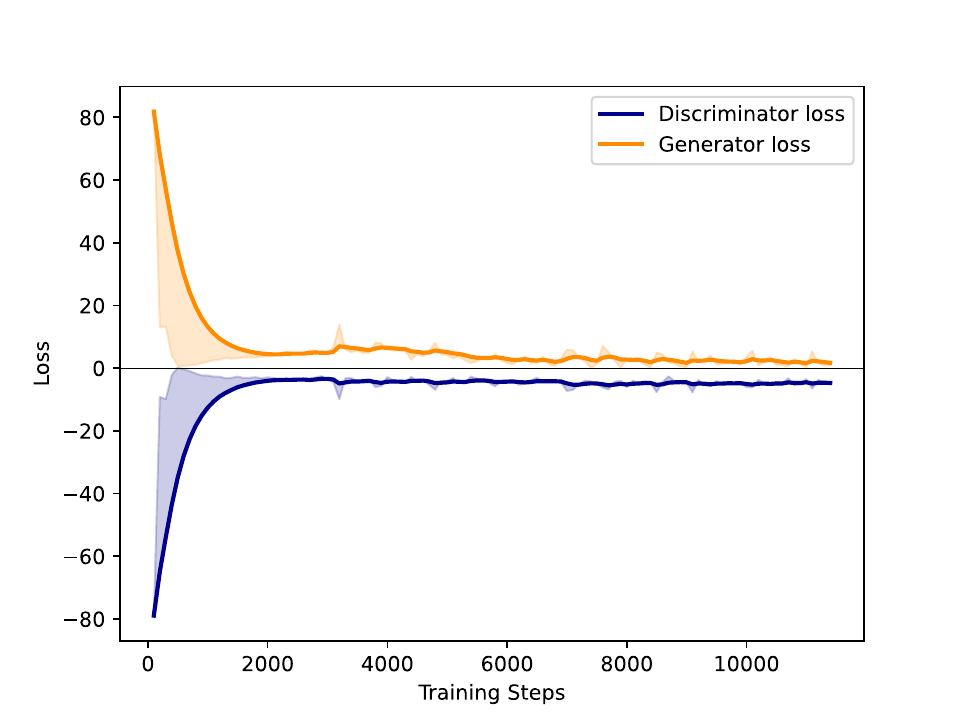}
\label{subfig:wgan_loss}
\caption{Comparative visualization of the discriminator and generator loss metrics over training iterations for a WGAN+GP model, showcasing the dynamic interplay and convergence patterns. The shaded area shows the maximum range of loss for individual training batches.}
\label{fig:wgan_losses}
\end{figure}

\begin{figure}[H]
\centering
\includegraphics[scale=0.8]{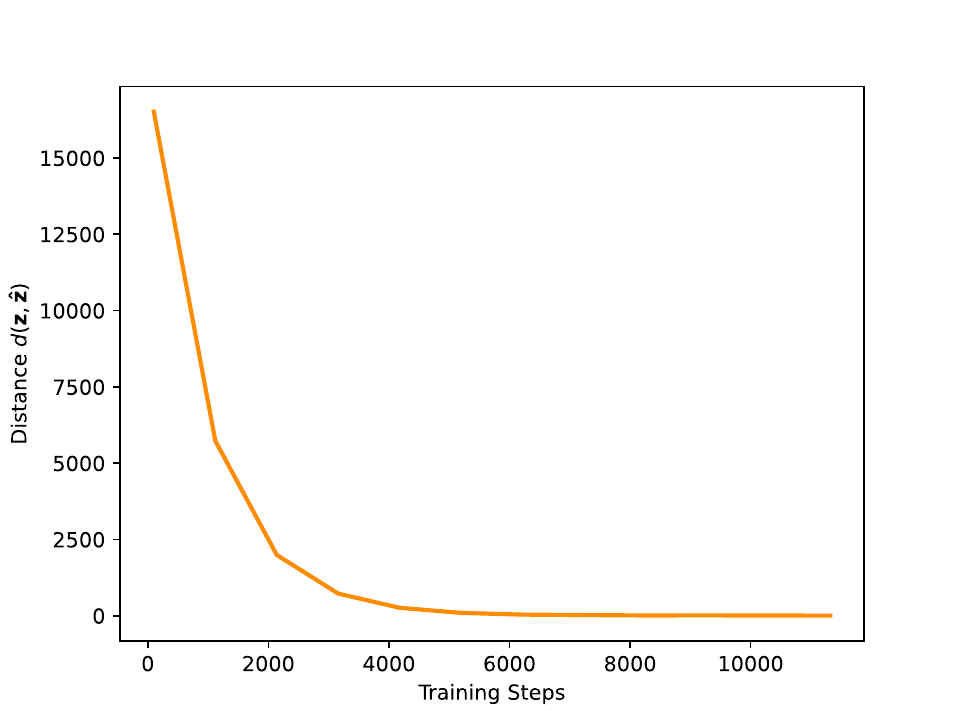}
\caption{Visualization of the distance metrics over $200$ epochs.The distance of the generated $\fz$ to the dataset $\mathcal{D}$ illustrates how closely the model-generated outputs match the dataset over training iterations.}
\label{fig:dist_loss}
\end{figure}

The WGAN+GP frameworks were utilized in our numerical evaluations to investigate their effectiveness in computing non-linear elastic materials through a data-driven approach. Figure \ref{fig:wgan_strain_stress} depicts the distribution of strain-stress achieved after $200$ training epochs. Due to the utilization of batch processing during training, the number of training steps exceeded this epoch count. The findings provide a profound understanding of the training quality and effectiveness.
\\
We investigated the loss values during training for a clearer perspective on model behavior. The  Wasserstein-enhanced architectures showcased robustness and consistency during training. Fig. \ref{fig:wgan_losses} displays the losses for the discriminator and generator of the model across the epochs. Notably, shallow loss values for either the generator or the discriminator can be counterproductive. It generally indicates that one network is dominating the other, leading to a stagnation in the training process. Ideally, there should be a balance where both networks challenge each other, encouraging continuous improvement.
Despite their effectiveness, traditional adversarial networks present non-interpretable loss values, making it challenging to discern training quality. The WGAN+GP approach offers direct insights into the quality of data generation, making it more user-friendly in solution analysis. In addition, we plot the minimum distance between the generated states and the data set $\mathcal{D}$ in Fig.~\ref{fig:dist_loss}. Given the data-driven approach, the learning process trains with a finite set of data points. Consequently, the losses do not converge to zero but to a positive lower bound. In our case, both losses decrease over training, representing this convergence.
In data-driven mechanics, the approach displayed a commendable ability to simulate stress-strain distributions. WGAN, with improved loss interpretability and smoother training, stands out as the preferable choice for intricate computational mechanics tasks. 

\section{Conclusion}\label{sec:conclusion}
The model-free data-driven method, developed by Kirchdoerfer and Ortiz, uses experimental data directly in simulations, bypassing the entire material modeling step. The paradigm uses nearest-neighbor clustering to reformulate boundary value problems. The approach has been diversified for many applications. Challenges such as data availability, noise, inconsistency, and high dimensionality frequently arise in the data-driven paradigm. Traditional analytical and computational methods may need to be adjusted when addressing these issues.
\\
Consequently, the incorporation of machine learning methods is considered, especially physics-informed neural networks. In solving boundary value problems with ANNs, the idea is to transform it into an optimization problem. The residual of the differential equations is minimized, and the neural network approximates the displacement and stress field. However, there are challenges with PINNs. There have been instances where the optimization yields solutions with unexpected or non-physical behaviors even when carefully tailored to encapsulate the physics. If we integrate the distance as an additional loss into the global loss, the whole problem becomes a nested optimization, leading to training challenges. In addition, approximated strain-stress fields can correspond to suboptimal data points influencing the direction and rate of the convergence. 
\\ 
To address these challenges, we consider the integration of PINNs with generative adversarial networks. GANs are proficient at generating outputs with the same properties as actual data, providing a potential approach to generating realistic strain-stress solutions. Their flexibility ensures adaptability across diverse data types suited for various physical conditions. Moreover, the inherent capability of GANs to distinguish and capitalize on intricate patterns may lead to a more robust representation of underlying physics. The combined PINN-GAN approach seeks to ensure physical consistency and alignment with observed data, leveraging the strengths of both methodologies. 
\\
This research introduced an approach to WGANs + GP tailored for data-driven mechanics problems. The generator is identified as a PINN, ensuring that generated outputs conform to underlying physical principles. Instead of random noise, the generator utilizes collocation points from the domain and maps them to neural network approximations of strain and stress fields. The discriminator is then trained using the generated and the closest actual strain-stress data. By integrating WGANs with physics-informed principles, the model outputs adhere to observed dataset statistics and differential equations. This results in improved optimization, more stable training, and accurate, physically consistent solutions. In this regard, we investigated a non-linear elastic plate with a hole benchmark. The results indicate that our proposed method provides reasonable outcomes. Furthermore, we observed robust and consistent training of the networks and noted the convergence of the data-driven solution as data size increased.
\\
As we advance our research, we aim to delve deeper into other convergence criteria for the GAN or WGAN. We plan to explore metrics such as the Inception Score \cite{salimans2016improved}, Frechet Inception Distance \cite{heusel2017gans}, and perceptual similarity measures \cite{zhang2018unreasonable} to provide a broader assessment of the generated outputs. These metrics will help to analyze the quality of the generated material states.
\\
Another area of interest is using the discriminator in the GAN framework for material identification. The discriminator's ability to distinguish between actual and generated outputs can be used to identify different material states. This approach could offer a novelty to classify materials, and we want to explore this further.
\\
In addition, we plan to extend our method to more complex and varied material properties. We also consider integrating advanced machine learning techniques to improve prediction accuracy, especially when dealing with sparse datasets. We are considering hybrid network architectures that combine convolutional and regression layers. The traditional image-based GAN structure inspires this design. By adding these layers, we hope to combine the advantages of image-based GANs with our current data-focused method.

\end{document}